\documentstyle[15lomcon,cite,url,graphicx,mathrsfs,lineno]{article}

\begin{document}

\title{PHOTON AND DI-PHOTON PRODUCTION AT ATLAS} 

\author{
  Marco Delmastro 
  \footnote{e-mail: \url{Marco.Delmastro@cern.ch}} 
  \footnote{On behalf of the ATLAS Collaboration}
 }

\address{CERN 
  \footnote{The author is now at LAPP (IN2P3/CNRS, France).}
}
 
\maketitle

\vspace{5mm}

\abstracts{ The latest ATLAS measurements of the cross section for the
  inclusive production of isolated prompt photons in $pp$ collisions
  at a centre-of-mass energy $\sqrt{s}$ = 7 TeV at the LHC are
  presented, as well as the measurement of the di-photon production
  cross section.
}

\newcommand{\ETg}   {E_{\mathrm{T}}^{\gamma}}
\newcommand{\mgg}   {m_{\gamma\gamma}}
\newcommand{\ptgg}  {p_{\mathrm{T},\gamma\gamma}}
\newcommand{\dphigg}{\Delta\phi_{\gamma\gamma}}
\newcommand{\ETiso} {E_{\mathrm{T}}^{\rm iso}}

\section{Overview}

The production of prompt photons at hadron colliders provides means
for testing perturbative QCD predictions, providing a colorless probe
of the hard scattering process. The dominant production mechanism of
single photons in $pp$ collisions at the Large Hadron Collider (LHC)
energies is $qg\to{}q\gamma$, while the production of di-photon final
states mainly occurs through quark-antiquark annihilation,
$q\bar{q}\to\gamma\gamma$, and gluon-gluon interaction
$gg\to\gamma\gamma$ mediated by a quark box diagram. In both single
and di-photon final states, parton fragmentation to photon also
contributes.

Because of the main production mechanism, the measurement of the
inclusive photon cross section at the LHC can constrain the gluon
density in protons. The study of the distribution of the azimuthal
separation between the two photons in di-photon events can provide
insight on the fragmentation model, while for balanced back-to-back
di-photons the di-photon cross section is sensitive to soft gluon
emission, which is not accurately described by fixed-order
perturbation theory. Di-photon production is also an irreducible
background for some new physics processes, such as the Higgs decay
into photon pairs.

We present here two measurements of the inclusive isolated prompt
photon production cross section as a function of the photon transverse
energy $\ETg$, using $pp$ collision data collected in 2010 with the
ATLAS detector \cite{ATLAS} at the LHC at a center-of-mass energy of 7
TeV. The former is based on an integrated luminosity $\int {\mathscr
  L} dt$~=~(0.88 $\pm$ 0.1) pb$^{-1}$
\cite{InclusivePhotons2010_880nb}, and provides a measurement of the
cross section for 15 $\leq\ETg<$ 100 GeV in the photon pseudorapidity
$\eta$ intervals [0,0.6), [0.6,1.37) and [1.52,1.81). The latter uses
the full 2010 data sample $\int {\mathscr L} dt$~=~(36.4 $\pm$ 1.2)
pb$^{-1}$ \cite{InclusivePhotons2010_35pb}, covering the 40
$\leq\ETg<$ 400 GeV $\ETg$ range and extending to the [1.81,2.37)
pseudorapidity region. 

We also present the measurement of the inclusive di-photon cross
section as a function of the di-photon invariant mass $\mgg$, the
di-photon system momentum $\ptgg$ and the azimuthal separation between
the two photons $\dphigg$, using an integrated luminosity $\int
{\mathscr L} dt$~=~(36.0 $\pm$ 0.1) pb$^{-1}$ \cite{DiPhotons2010}.

\section{Photon selection, reconstruction and identification in ATLAS}

Single photon events are triggered in ATLAS using a high-level trigger
with a nominal transverse energy threshold of 10 GeV
\cite{InclusivePhotons2010_880nb} or 40 GeV
\cite{InclusivePhotons2010_35pb}; di-photon events are triggered by
two photon candidates having $\ETg>$ 15 GeV
\cite{DiPhotons2010}. Using unbiased or lower-threshold triggers,
these triggers are found to be fully efficient for photons and
di-photons passing the selection criteria of the analyses.

Photon candidates depositing their energy in the ATLAS Liquid Argon
(LAr) electromagnetic calorimeter (EMC) in the regions $|\eta|<$1.37
and 1.52$\leq|\eta|<$2.37 are reconstructed. The photons converting in
$e^+e^-$ pairs before reaching the EMC (about 30\% in the samples
under study) are separated from electrons by associating the
reconstructed tracks and conversion vertices to the EMC energy
deposit. The overall photon reconstruction efficiency is about 85\%
(75\%) for $|\eta|<$1.37 (1.52$\leq|\eta|<$2.37), the main losses
being due to nonoperational LAr EMC readout modules during the 2010
data taking. In the inclusive photon analyses photon candidates are
required to have reconstructed $\ETg$ larger than 15 GeV
\cite{InclusivePhotons2010_880nb} and 45 GeV
\cite{InclusivePhotons2010_35pb}; in the di-photon measurement both
photon candidates in a event must have $\ETg>$ 16 GeV.

Background from non-prompt photons originating from decays of leading
neutral mesons inside jets (e.g. $\pi^0$) is suppressed by means of
selections on the electromagnetic shower momenta, and of a requirement
on the photon isolation in the EMC. Photon candidates must pass tight
identification criteria based on nine discriminating variables
computed from the lateral and longitudinal profiles of the energy
deposited in the EMC, and in the hadronic calorimeter behind it. The
first LAr EMC layer is finely segmented such to allow the resolution
of two maxima in the energy deposit, typical of the superposition of
two photons from a neutral meson decay. The efficiency of these
selections ranges from $\sim$ 60\% to $\sim$ 90\% for increasing
$\ETg$.

The ￼photon transverse isolation energy $\ETiso$ is computed from the
sum of the energies in the LAr EMC cells in a cone of radius 0.4 in
the $\eta-\phi$ plane around the photon candidate axis. The
contribution to $\ETiso$ from the photon itself is subtracted, as well
as the energy from the soft-jet activity from the underlying event and
from event pileup \cite{UEcorrection}. All measurements presented here
require $\ETiso<$ 3 GeV.

\section{Background subtraction}

After the selection criteria described above, a residual contribution
of background candidates pollutes the photon and diphoton samples.
In the inclusive photon analysis, this background contamination is
estimated using a data-driven counting technique based on the observed
number of events in the control regions (sidebands) of a
two-dimensional plane formed by the photon transverse isolation energy
and a photon identification variable.
Corrections for signal leakage in the background control regions and
for correlation between the two variables in background events are
taken into account.

The same sideband method is iterated on di-photon event on the leading
and sub-leading photon candidates, allowing to separate the di-photon
signal from the photon--jet and jet--jet background
components. Alternatively, a matrix approach classifying the events in
categories according to whether each of the tight photon candidates
passes or not the isolation criteria, or a template approach fitting
the distributions of the isolation energy profiles, are used and give
compatible yield results.

In both inclusive and di-photon analyses the residual contamination
from electron misidentified as photons is evaluated and subtracted.

\section{Cross section measurements}

\begin{figure}[t]
  \centering
  \begin{minipage}[c]{1.2\linewidth}
    \hspace{-0.07\textwidth}
    \includegraphics[width=0.24\textwidth]{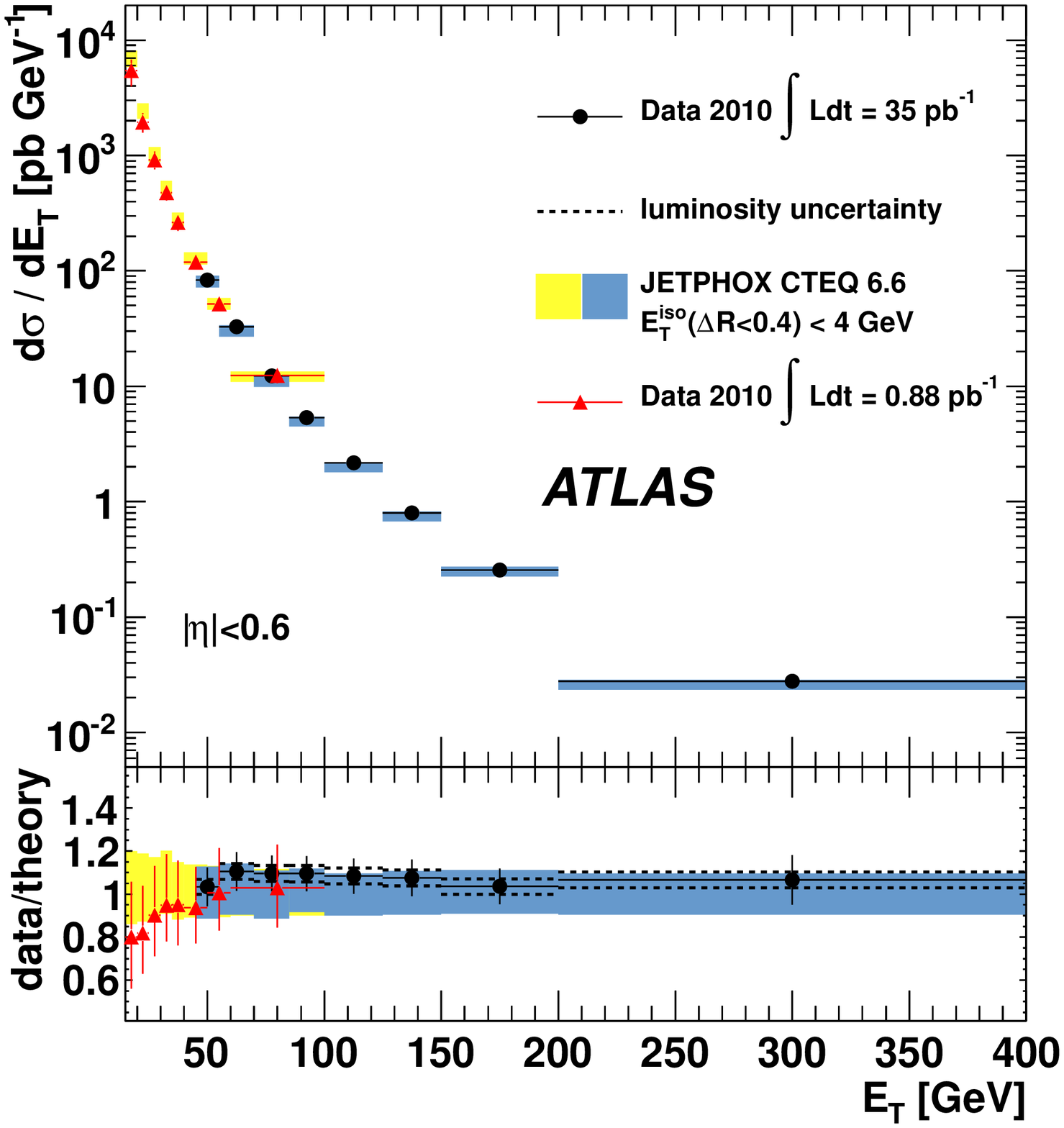} 
    \hspace{-0.02\textwidth}
    \includegraphics[width=0.24\textwidth]{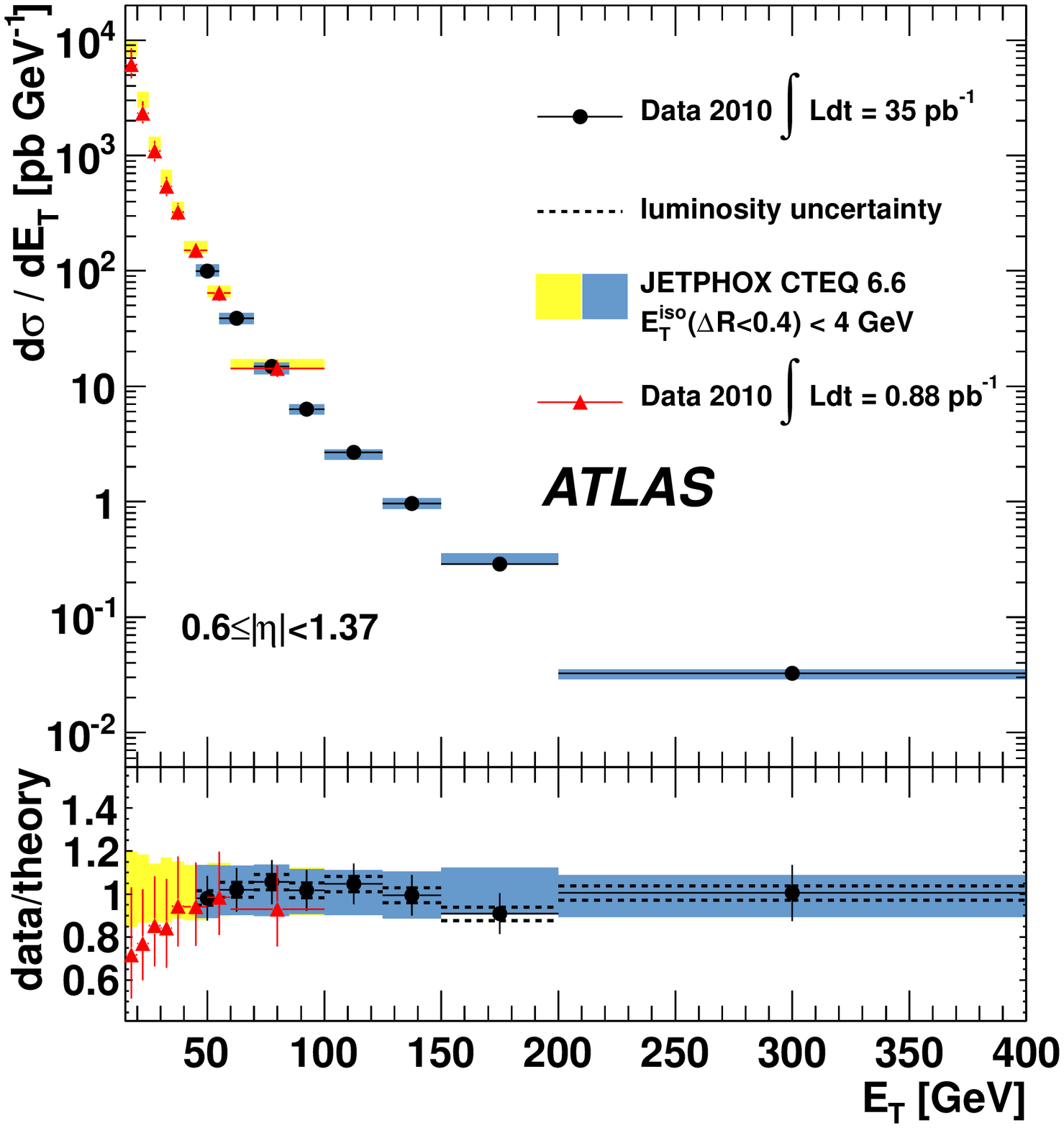}  
    \hspace{-0.02\textwidth}
    \includegraphics[width=0.24\textwidth]{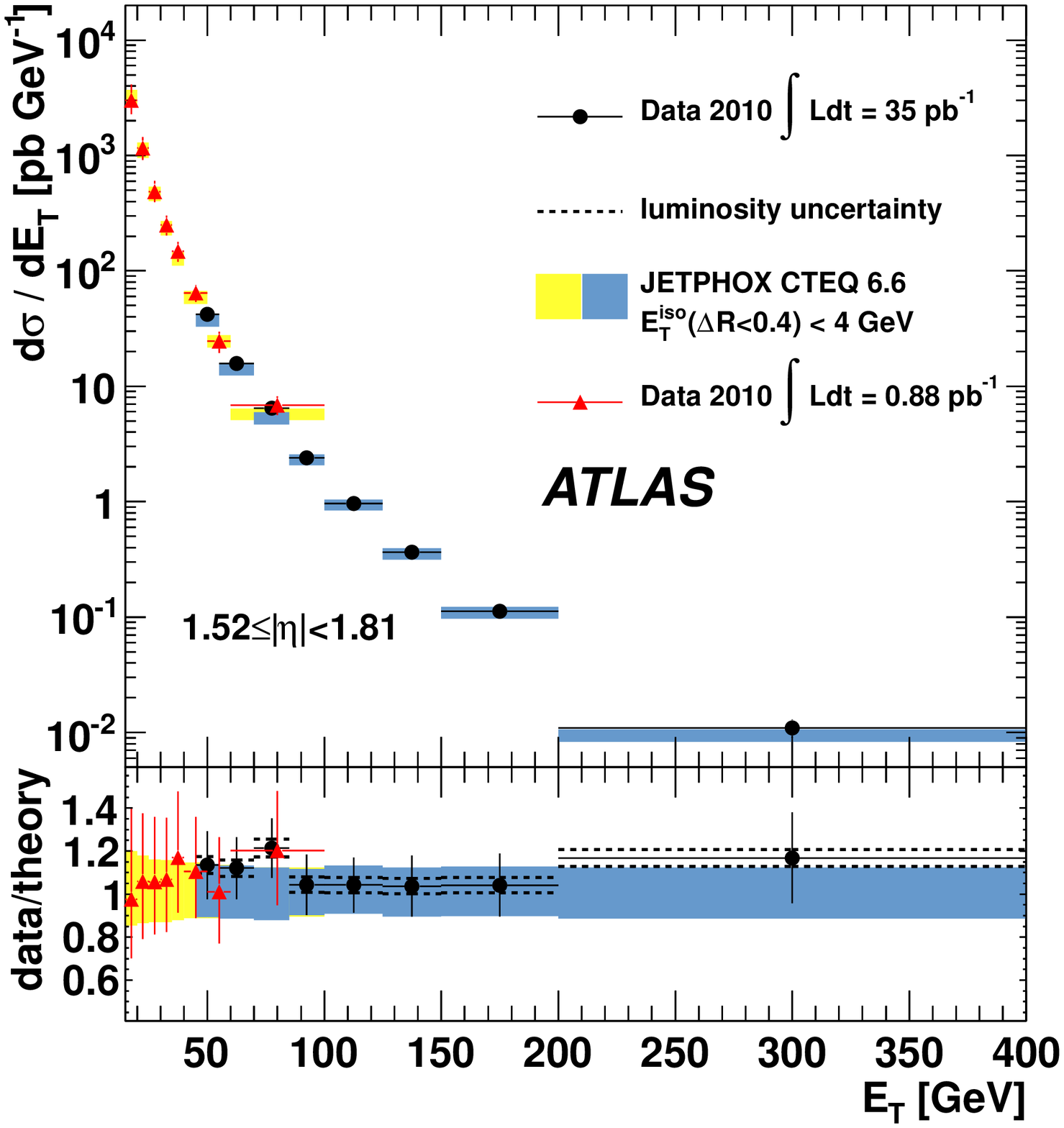}
    \hspace{-0.02\textwidth}
    \includegraphics[width=0.24\textwidth]{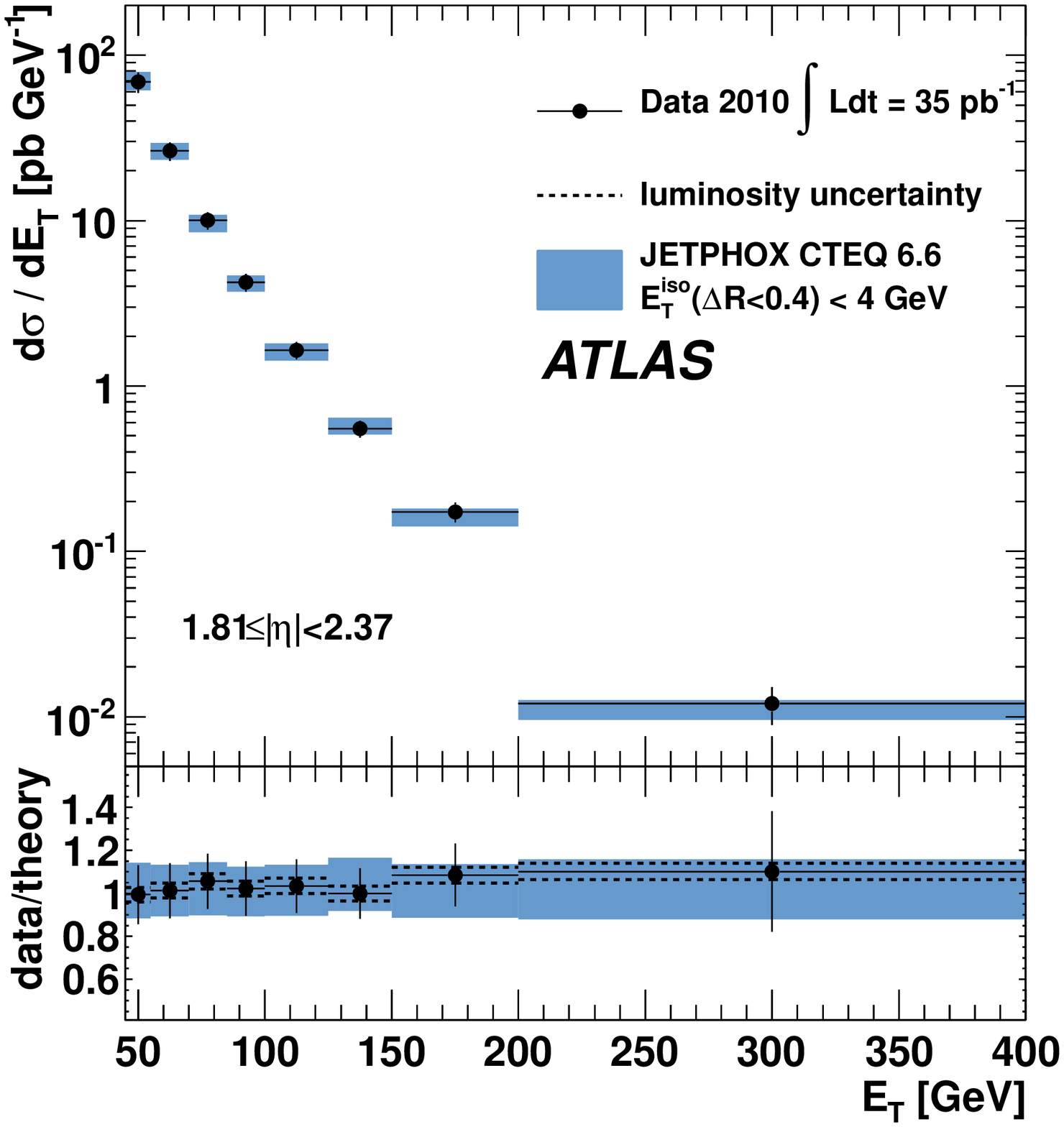} 
  \end{minipage}
  \vspace{-3mm}
  \caption{Measured cross section of isolated prompt-photon production
    as a function of $\ETg$ in different pseudorapidity ranges,
    compared with theoretical predictions. The red triangle represent
    the results from \cite{InclusivePhotons2010_880nb}, the black dots
    those from \cite{InclusivePhotons2010_35pb}.}
  \vspace{-3mm}
  \label{fig:inclusive_cross_section}
\end{figure}

Inclusive prompt photon and di-photon cross sections are respectively
measured in different bins of $\ETg$ and $\mgg, \ptgg, \dphigg$ from
the extracted signal yields, the corresponding integrated
luminosities, and the trigger, reconstruction and selection
efficiencies. The measured cross sections are affected by different
systematic uncertainties, primarily associated to the uncertainty on
the photon reconstruction efficiency (3-4\% due to the isolation
efficiency cut, and 1-2.5\% associated to the limited knowledge of the
material upstream the EMC); to the uncertainty on the photon
identification efficiency (ranging from 8\% to 1.5\%, the higher
values being applicable at lower $\ETg$); to uncertainty on the signal
yields due the background subtraction technique (at most 10\%, mostly
associated to the definition of the background control regions and the
photon energy scale).

Figure~\ref{fig:inclusive_cross_section} shows the inclusive photon
cross sections as a function of $\ETg$ in four different $\eta$
regions. The theoretical pQCD cross sections, computed with a
fixed-order NLO parton-level generator (JETPHOX \cite{JETPHOX}) for
photons having parton transverse energy in a cone of radius 0.4 around
the photon smaller then 4 GeV, are overlaid (yellow and blue bands,
accounting for the scale and PDF uncertainties). The measured cross
sections are in good agreement with the theoretical predictions for
$\ETg>$ 35 GeV, while at lower $\ETg$, where the contribution from
parton-to-photon fragmentation is larger, the theory tends to
overestimate the data, possibly hinting to the need of more accurate
predictions.

\begin{figure}[t]
  \centering
  \includegraphics[width=0.32\textwidth]{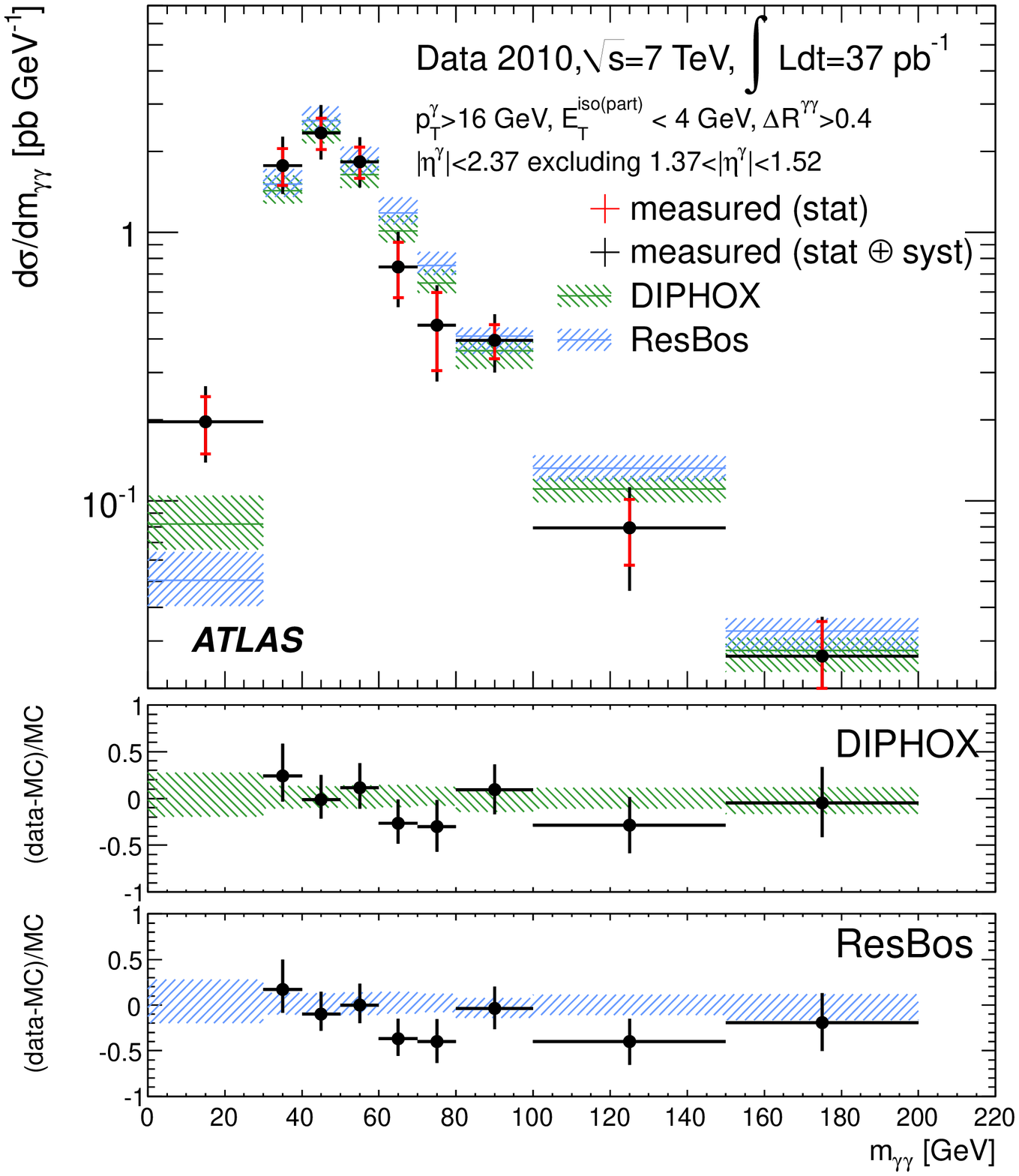}
  \includegraphics[width=0.32\textwidth]{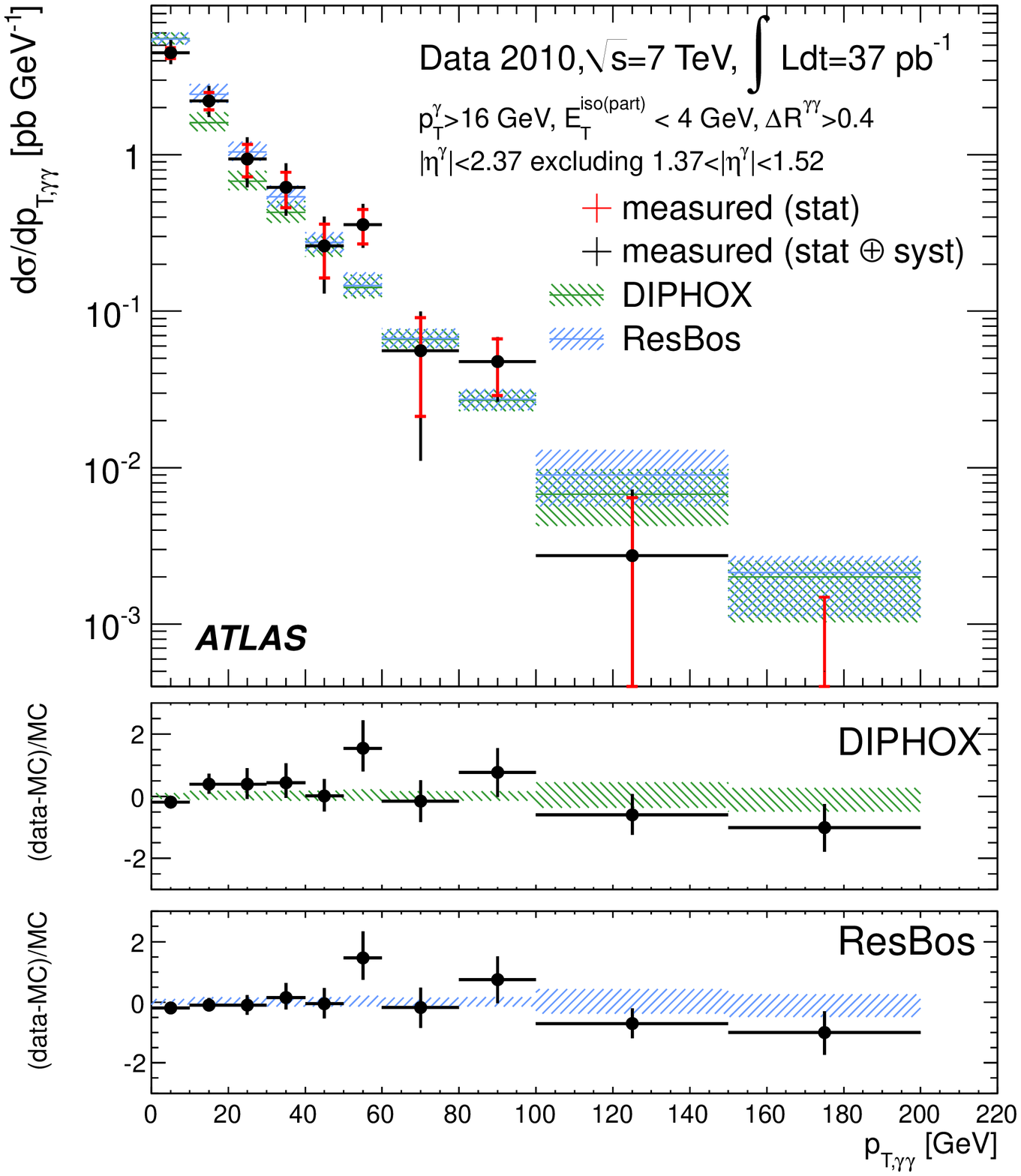}
  \includegraphics[width=0.32\textwidth]{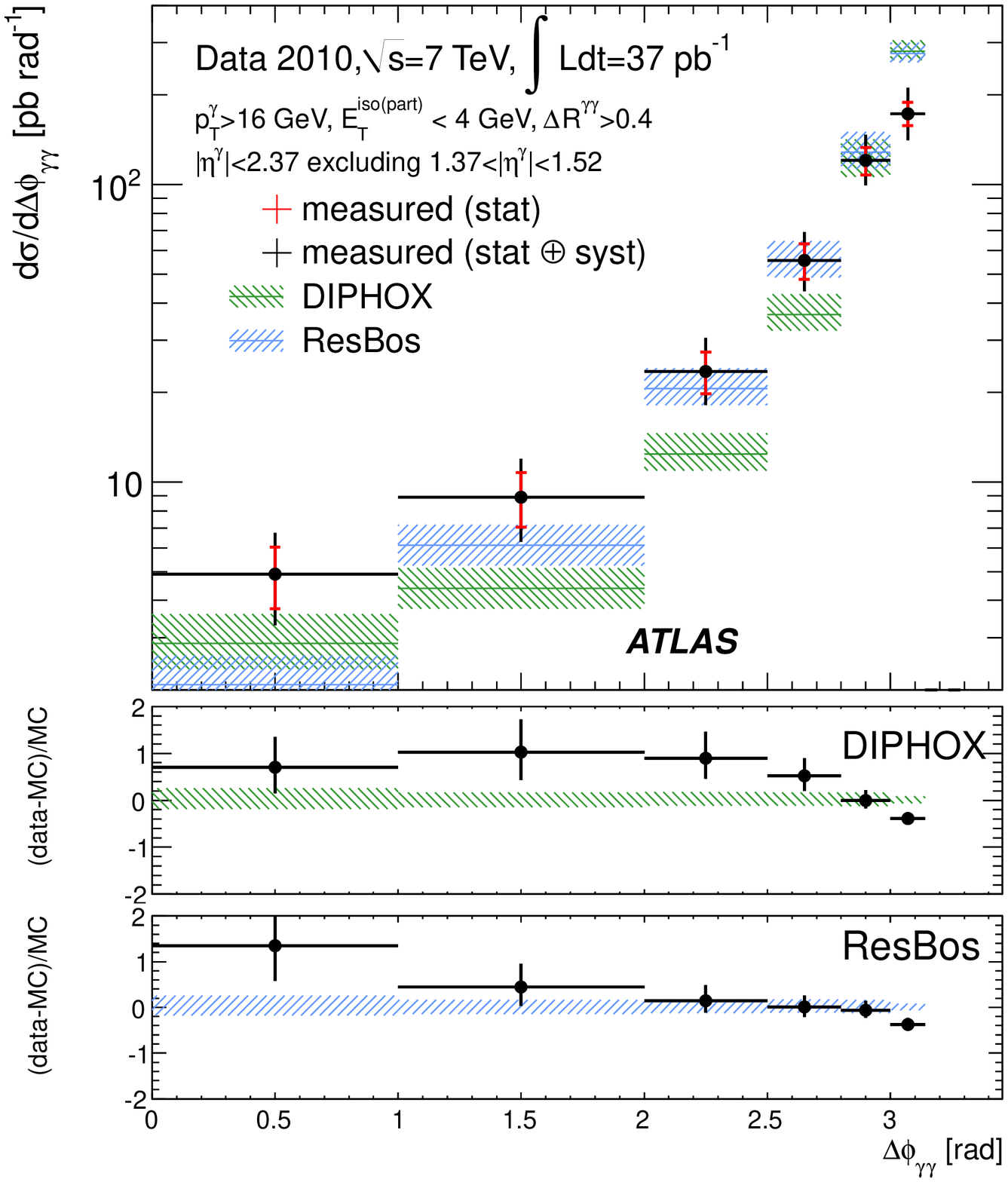}
  \vspace{-3mm}
  \caption{Measured cross section of isolated di-photon production as
    a function of $\mgg, \ptgg$ and $\dphigg$, compared with
    theoretical predictions.}
  \vspace{-3mm}
  \label{fig:diphoton_cross_section}
\end{figure}

Figure~\ref{fig:diphoton_cross_section} shows the diphoton cross
section as a function of $\mgg$, $\ptgg$ and $\dphigg$. Two
theoretical predictions are overlaid for photons having parton
transverse energy in a cone of radius 0.4 around the photon smaller
then 4 GeV, one corresponding to a fixed-order NLO parton-level
generator calculation (DIPHOX \cite{DIPHOX}), the other featuring
transverse momentum resummation (RESBOS \cite{RESBOS}). The agreement
is generally good, but some deviations are observed for low $\dphigg$
values, where both theoretical predictions underestimate the
measurements. In this region the LO elements do not contribute to the
cross section, and NLO is the first order giving non-zero
contributions: more accurate NNLO predictions would provide
clarifications on these residual discrepancies.

\section*{References}

\bibliographystyle{unsrt}

\begin{thebibliography}{99}
{\footnotesize
\bibitem{InclusivePhotons2010_880nb} ATLAS, 
  Phys. Rev. {\bf D83} (2011) 052005 [arXiv:1012.4389].

\bibitem{InclusivePhotons2010_35pb} ATLAS, 
  submitted to Phys. Lett. B. [arXiv:1108.0253]. 

\bibitem{DiPhotons2010} ATLAS, 
  accepted for publication by Phys. Rev. D  [arXiv:1107.0581].

\bibitem{ATLAS} ATLAS, 
  JINST 3 (2008) S08003.
  ATLAS, 
  [arXiv:0901.0512].

\bibitem{UEcorrection} M.~Cacciari, G.P.~Salam and S.~Sapeta,
 JHEP04, 065 (2010). 

\bibitem{JETPHOX}  M.~Fontannaz, J.P.~Guillet and G.~Heinrich,
  Eur. Phys. J. {\bf C21} (2001) 303--312. 

\bibitem{DIPHOX} T. Binoth, J. P. Guillet, E. Pilon, M. Werlen, 
  Eur. Phys. J. {\bf C16} (2000) 311--330. 

\bibitem{RESBOS} C. Balazs, E. L. Berger, P. M. Nadolsky, C. -P. Yuan, 
  Phys. Rev. {\bf D76} (2007) 013009. 
}￼
\end{thebibliography}

\end{document}